# Investigation of the anatase-to-rutile transition for TiO2 sol-gel coatings with refractive index up to 2.7

Martin O'Byrne,[a] Badre Kerzabi,[b] Marco Abbarchi,[a,b] Alejo Lifschitz,[c] Tony Zamora,[c] Victor Malgras,[a] Anthony Gourdin,[b] Mehrnaz Modaresialam,[b] David Grosso,*[a,b] Magali Putero*[d]

[a]CINaM, Aix-Marseille Université, Marseille, 13288, France
[b]Solnil, 95 Rue de la République, Marseille, 13002, France
[c]Reality Labs, Meta, 9845 Willows Rd. NE, Redmond, WA, USA
[d]IM2NP, Aix Marseille Université, Marseille, 13288, France

## Abstract

This work describes the elaboration of rutile titanium dioxide films with high refractive indices and low scattering by sol-gel process and controlled crystallization. The evolutions of the optical properties and crystalline structure of sol-gel processed titania coatings on fused silica were investigated for different thermal budgets of the annealing post-treatment using ellipsometry, spectrophotometry, X-ray diffraction and electronic microscopy. It reveals that anatase and rutile coatings with refractive indices of 2.5 and 2.7 can be prepared with associated optical loss of 0.5% and 1%, respectively, which are excellent compromise for applications in integrated photonics. These evolutions are associated to the thermally induced mass transfer and phase transitions occurring during thermal annealing that involves first the nucleation growth and sintering of anatase polyoriented nanocrystals, followed by the transformation into rutile polyoriented nanocrystals. Concomitantly, rutile crystals with (110) faces parallel to the surface consume surrounding anatase and rutile nanocrystals by diffusive sintering to yield micron-size rutile monocrystalline and monooriented platelets patchwork, exhibiting refractive index of 2.73 and 1.2% optical loss. The formation of these platelets is governed by surface energies and is responsible for the increase in optical loss.

## 1. Introduction

The development of integrated photonics technologies for augmented reality[1], virtual reality[1,2], or multispectral sensors[3–5], relies on the use of dielectric and optically passive materials, with high refractive index and low absorption and scattering. Based on their processability in conventional integrated circuits foundries, the available candidates are mainly silicon and silicon nitride[6–8]. The extinction coefficient of Si ranges from 6 to $10^{-3}$ between the near UV and the near IR, making it a dielectric material of choice for near IR applications[9]. Being lossy in the visible range, Si can be replaced by $Si_3N_4$, whose extinction coefficient is almost zero, with a refractive index that peaks at 2.15 in the near UV[9]. However, metal oxide materials with high refractive index that are transparent in the visible range (e.g. $TiO_2$, $HfO_2$, $Nb_2O_5$, $ZrO_2$)[10–12] exist and have the advantage of being cost-effective and easily processed as thin films in presence of $O_2$, such as in air by sol-gel liquid deposition, or by vacuum deposition such as CVD. Titanium oxide in its most stable form, rutile, is undoubtedly the best candidate as its refractive index can reach between 2.6 and 2.9 due to its birefringence[13,14], which places it at the top of the list of transparent oxides with high optical density. It also has the advantage of being extremely stable as it is resistant to chemical attack, abrasion, and temperature variations[15], not to mention its natural earth crust-abundance and biocompatibility.

However, these ideal performances of titanium oxide[16] are rarely obtained experimentally owing to defects in the arrangement of the material, occurring during deposition or post-processing densification[17]. Both in sol-gel[18] or vacuum[19] deposition processes, densification of the material requires an increase of the mobility of the atoms by high temperature annealing so that they can relax in an energetically favorable stable state. Thus, the optimal densification state is usually achieved through heat treatment-induced diffusive sintering. At standard atmospheric pressure, the titanium oxide deposited by these techniques is either amorphous, or in its metastable crystalline form anatase, or in its thermodynamically stable crystalline form rutile [20]. In theory, the refractive index increases as $TiO_2$ transits from the amorphous to the rutile phase. In practice, the measured index will be governed by the grain size, the presence of grain boundaries, the presence of inter-grain porosity and the volume fraction of the phases[17]. Thus, to obtain a $TiO_2$ layer with a maximum refractive index, this layer should consist only of rutile, and contain a minimal number of pores and grain boundaries, as can be expected from large crystals extending longitudinally on the surface and with a compact arrangement. On the other hand, the crystal size should remain below the visible range (< 300 nm), beyond which light scattering shall induce unwanted losses and reduce photonic device efficiency. So far, the control over high high-index and low low-scattering has been elusive and only relatively low-n titania has been used for photonic devices[21,22].

anatase, with a tetragonal structure, is the most studied phase because of its intrinsic semi-conducting properties rendering a material of choice for various photocatalytic reactions[23]. rutile, also with a tetragonal structure but with a denser mesh, is less chemically active, partly due to its indirect and larger gap and lower charge carrier mobility. It is less used for photocatalysis applications[24], but will be preferred for optics and photonics requiring higher refractive indices[25]. The superior photocatalytic activity of anatase is also associated with the surface energy of its crystal facets being lower than that of rutile crystal facets. This property also favors the stabilization of small size crystals (10 - 50 nm), thus allowing for the development of a larger specific surface area[26]. This is why the transformation of anatase $TiO_2$ into small domains of rutile is difficult to control. Different reports, correlating the effects of chemical and process parameters (precursors, dopants, media, thermal treatment, etc.) revealed that the transition of the kinetically stabilized anatase phase into a thermodynamically stable rutile phase can be obtained under very different conditions depending on the systems[27–29]. Under rapid nucleation conditions, the formation of small densely packed anatase particles is dominant. The subsequent transition into rutile is less favoured due to a surface energy three-times higher[30] and, therefore, an intermediate stage is necessary, where the particles will grow beyond a critical size, enabling the stability of the rutile structure[31]. The size, morphology and arrangement of anatase crystals on the substrate surface are thus critical parameters that will influence the organization of rutile grains after the transition[26,28]. Since the formation of rutile crystals from anatase is thermodynamically favorable, due to a lower Gibbs free-energy between the two phases (transformation enthalpy of anatase to ~~Rutile~~ rutile: $-2 > \Delta G_{298K} > -12$ kJ.mol$^{-1}$)[26], the difficulty in lowering the scattering by grains, consists of maintaining the rutile crystals as small as those formed in anatase phase. Unfortunately, the threshold domain size obtained during the intermediate anatase growth falls in the visible light scattering range, so the formation of both a high-refractive index and low-diffusion dense rutile thin film may not be easily achieved. The formation of dense rutile $TiO_2$ layers with high index and minimal defects represents a challenge yet to be overcome, since it is limited by the difficulty to pack low-dimensional crystalline nano-objects without residual inter-particle porosity. Motivated by the latent potential of this material, intensive investigations have been carried out on this crystal phase transition. Note that sol-gel deposition on solid substrates impose an important kinetic component to the thin film formation, therefore rules dictating thermodynamic relaxation will not necessarily apply, making the whole process difficult to model. Therefore, despite the long history of research on $TiO_2$, there is still a wide knowledge gap related to crystal transition and growth mechanisms associated to thin films.

This work is a study of the anatase-rutile transition at 1100 °C of 50 nm thick $TiO_2$ films prepared by sol-gel process on fused silica, monitoring and correlating their optical and crystallographic evolution. We show that low-index coatings are mainly composed of anatase while the high index ones are composed of large rutile crystals in the form of platelets whose thickness corresponds to the thickness of the film, fitted to each other and preferentially oriented with the [110] crystallographic direction perpendicular to the substrate surface. The measured optical properties show that a combination of low scattering, between 0.8 and 1.6% and high refractive indices, 2.52 and 2.73 (at 520 nm), respectively, is possible.

## 2. Experimental section

The sol-gel solutions were prepared by dissolving 1M Titanium Tetraisopropoxide (TTIP - Technipur™, for synthesis) in ethanol, containing 1.33 M water and 0.8 M $HNO_3$. The solution is then left to stir at room temperature for at least 1 h for stabilization, it remains stable and can be used for up to 48 h without impacting the quality of the samples. The films are prepared by dip-coating the cleaned fused silica substrates under dry conditions in the solution with a withdrawal speed of 1.6 mm $s^{-1}$. The films are then pre-stabilized at 250 °C (temperature lower than the crystallization temperature of anatase) for 1h. Following the pre-stabilization, the films are fired by being placed in a tube furnace pre-heated at 1100 °C in air. They are then maintained in the said furnace for 60 s + 5 to 60 s. The mandatory 60 s corresponds to the time required for the sample to thermalize and has been determined by measuring the temperature of a bare substrate with a thermocouple. This measurement has been repeated 5 times and show a variation of ± 3 s depending on the position of the substrate and the thermocouple.

The films, with thicknesses around 50 nm, were observed by electronic microscopy: SEM samples were coated with a 2 nm Ir layer using a Quorum 150T S Plus sputter coater and imaged with a Zeiss GeminiSEM 500 operating at 3 kV acceleration voltage and a working distance of 3 mm. PED-TEM samples were mechanically polished, followed by ion broad beam milling, resulting in the removal of ~10 nm of material. PED data for the top, polished surface was collected using a Nanomegas ASTAR system coupled to a ThermoFisher Scientific FEI Tecnai TF-30 TEM. $TiO_2$ rutile and anatase were used as templates for orientation mapping. The minimal grain size threshold was set to 25 nm and an index of 300 was used as the cut off threshold for any amorphous areas). X-ray diffraction patterns were acquired on a conventional diffractometer (PANalytical Empyrean) using Cu radiation (λ = 0.154 nm) and a rapid detector (PANalytical PIXcel) in Bragg–Brentano geometry. The average grain sizes were determined by fitting the Bragg peaks, using the integral breadths of the diffraction lines[32], considering the experimental integral breadths, and neglecting microstrain contribution. The full-profile-fitting refinements were also carried out by the Rietveld method, using the Profex software[33] to deduce the percentage of each phase. Spectroscopic measurements were performed using a spectrophotometer (from Perkin Elmer) mounting an integrating sphere. Spectroscopic ellipsometry analysis was performed on a Woollam M2000 V using a "GenOsc" model of the Woollam "Complete EASE" software in which the absorption associated with the band gap was modeled by a Tauc-Lorentz type Oscillator and by fixing a roughness of 5 nm, typical for this type of sample.

## 3. Results and discussions

Figure 1 shows the evolution of the refractive index of the layers as a function of residence time at 1100 °C. The index increases with the heat budget, from 2.52 to 2.73. The ellipsometry fits of these layers have Mean Square Errors (MSE) between 2.5 and 8 which reflects a good correlation between the experimental values and the simulation and thus an excellent optical quality of the layers, especially for the lowest indices. The n, k dispersions provided in Figure 1B and C reveal distributions typical of a dielectric material with k = 0 in the visible for the samples heated for 5, 10 and 20 s. Absorption in the near UV, however, increases with treatment time. Samples treated with lower thermal budgets led to films with lower indices, mainly composed of anatase, that have not been studied in details in this work.

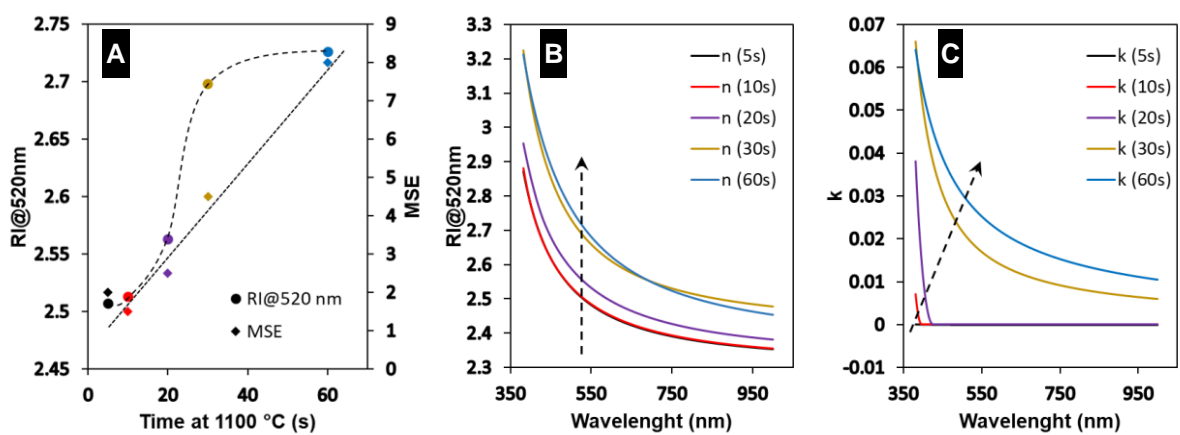

Figure 1. A: Refractive index at 520 nm measured by spectroscopic ellipsometry using GenOsc (Tauc-Lorentz) model, and corresponding Mean Square Error (dashed lines are here as eyes guides) for TiO₂ samples heat-treated at various durations. B and C: corresponding n and k dispersions.

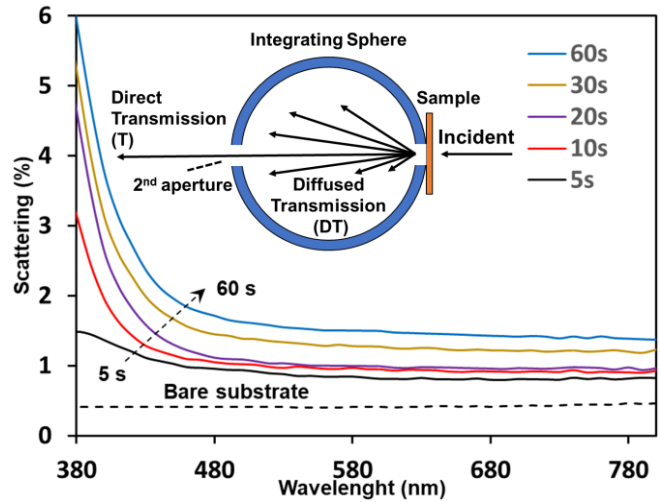

Figure 2. Scattering spectra of films treated for different durations at 1100 °C, compared to the bare substrate. Scattering (S) has been plotted as the ratio between the diffused transmission (DT) obtained with an integrating sphere equipped with a 2° angle aperture to eliminate contribution of the specular transmission, and the Total transmission (TT) obtained with the obstructed aperture (see inset).

The ellipsometric study was complemented by spectroscopic transmission analyses of these films (Figure 2) to determine the scattering losses that are a recurrent problem in high index materials for optical and photonic applications. The scattering spectra (S), which also encompasses the intrinsic absorbance of the films, shows a progressive increase in the loss (S = measured S - substrate S) from 0.4 to 1.2% for 5 to 60 s of treatment. The increase is greater between 20 and 30 s and also corresponds to the greatest increase in refractive index (Figure 1 A), but also to a significant change in the organization of the material as we will show in what follows. This increase in scattering is most certainly responsible for the increase in k measured by ellipsometry. $TiO_2$ is not supposed to absorb in the visible, however the ellipsometer cannot model scattering, therefore the reflective losses are wrongly associated to an increase in absorption (k). The slightly higher MSE for the last points is somewhat an indication that the model diverges a little because of such scattering.

Figure 3 shows the evolution of the grain texture of the layers as a function of the calcination time through SEM observation. No major defects larger than a few hundred nanometers were observed confirming the good optical quality of these layers. Below 30 s annealing time, the surfaces and sections show a compact stack of relatively homogeneous nanosized grains, whereas above 30 s, the grains become more visible and coexist with sharp fractures delimiting micron-sized platelets along with cavities or pores (black spots) homogeneously dispersed on the surface. The sections also show that at 60 s of treatment, the film is entirely composed of randomly shaped platelets arranged in a dense fashion. In addition, the pores have also disappeared. The lower amount of grain boundaries (Figure 2F) implies that each domain consists of a single monocrystalline platelet. This investigation reveals that the $TiO_2$ film undergoes a considerable transformation of its texture between 20 and 30 s of treatment at 1100 °C, causing an increase of both the refractive index (Figure 1) and the loss by diffusion (Figure 2). Indeed, considering that the difference in absorption between the films treated for different durations should be negligible, the increase in scattering loss is mainly due to the structural transformations observed by SEM. Important information regarding the evolution of the material can be deduced from the X-ray diffraction (Figure 4). The XRD patterns obtained for each film show that the transition of the anatase into rutile occurs gradually between 5 and 30 s. According to Rietveld's refinements and Scherrer's modelling, the layer is composed of 5 % rutile at 5 s and 100 % above 30 s, while the grains' size does not vary significantly even after 30 s of heat treatment. The Bragg-Brentano geometry of the X-ray setup in use allows us to observe only the crystal planes parallel to the substrate surface. Thus, the obtained grain size corresponds to the grain dimension orthogonal to the surface. This dimension is similar to the film thickness measured through ellipsometry and is consistent with the single crystal nature of the platelets observed in Figure 3, constrained into 2D confinement.

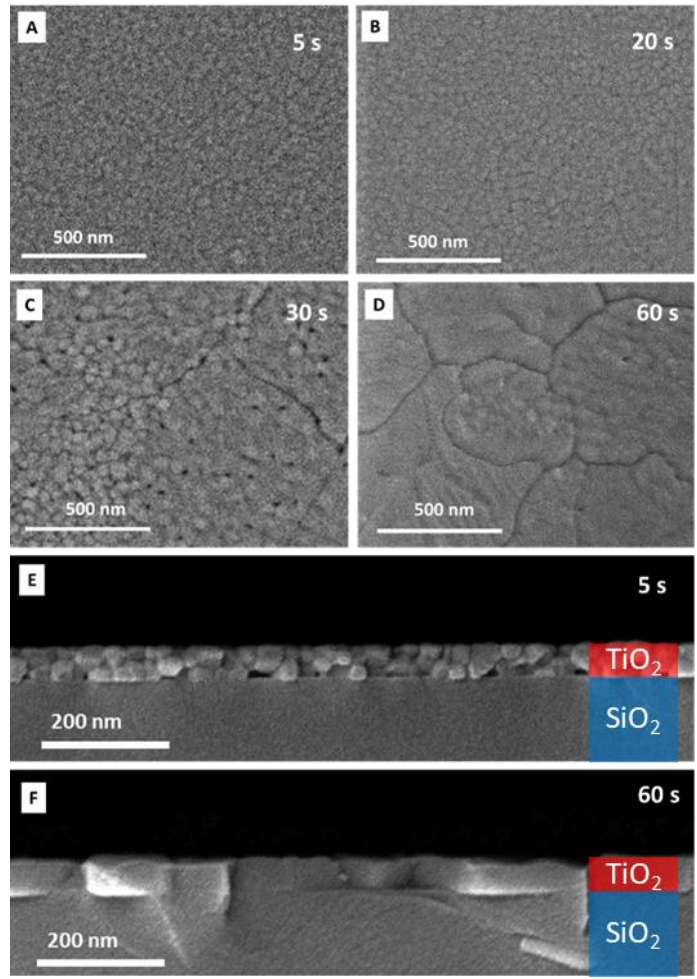

Figure 3. SEM images of coatings treated for different durations at 1100 °C: A-D: top-views. E,F: cross-sectional views. The low resolution is due to the non-conductive fused silica substrate.

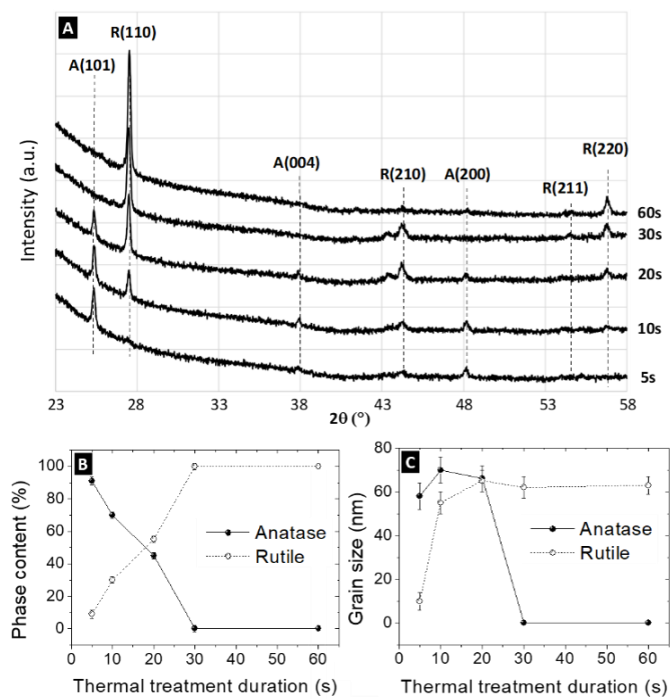

Figure 2. XRD (Bragg-Brentano) analysis of films treated different times at 1100°C. A: patterns with corresponding Miller indices indexed for anatase (A) and rutile (R) phases; B and C: phase proportion and average grain size deduced from Rietveld refining and Scherrer formula, respectively.

If we now compare the patterns at 30 and 60 s, we observe a clear decrease of the (210) and (211) reflections of rutile to leave mainly the (110) reflection. This clearly indicates that the nano-crystallized material composed of poly-oriented anatase undergoes a first transformation into poly-oriented rutile grains before reorienting and sintering simultaneously into micron-sized platelets with the (110) facets oriented preferentially parallel to the surface. Note that the substrate being amorphous fused silica, no epitaxial relationship is expected, and the reorientation is therefore mainly governed by the surface and interface energies.

The 20 s processed sample was analyzed by TEM-PED to better visualize this transformation. Figure 5 combines an orientation and a phase mapping images for this intermediate sample. Two very distinct areas can be seen on the right and left of both images. The area on the left shows dense patches of rutile (red) of the same orientation whereas the area on the right shows a randomly poly-oriented distribution of anatase grains of a few tens of nanometers. The right side of the phase image also shows the presence of isolated and randomly orientated rutile crystals among the anatase particles. The anatase-to-rutile transition thus occurs at the interface of those two zones (plate/nanocrystals), suggesting a sintering mechanism of rutile crystals with the (110) orientation and poly-oriented anatase crystals. Therefore, the thermal treatment of a TiO₂ sol-gel layer at 1100 °C leads to the crystallization into rutile by passing through the anatase phase, as expected.

From the data shown in Figures 3, 4 and 5, the evolution of the film crystal transition can be assumed to follow the steps illustrated in Figure 6. First, a homogeneous nucleation into poly-oriented anatase nanoparticles (NPs) occurs below 5 s annealing. This nucleation is followed by a diffusive, sintering-driven growth, allowing the particles to reach transverse dimensions close to the film thickness, resulting from the spatial confinement between both interfaces. The poly-oriented anatase particles formed on the amorphous fused silica substrate are stabilized by their low global energy (surface + volume) associated with the low surface energy of the anatase phase[34]. However, an increasing fraction of these particles transforms into rutile during this period. After 10 s, 30 % of the material is in the form of poly-oriented rutile NPs and randomly dispersed within the initial network of poly-oriented anatase. It has been demonstrated experimentally (by HRTEM) as well as by DFT simulation (stochastic surface walking theory) that the transformation occurs by reconstruction of the (112) face of anatase into the (100) face of rutile by passing through a thin layer of brookite[35], resembling a nucleation growth where the (100) face of rutile propagates within the anatase crystal. Therefore, there are no reasons for the rutile crystals to adopt a different orientation with respect to that of the starting anatase crystals, which results in the formation of poly-oriented rutile grains in presence of anatase grains. At 20 s, the proportion of rutile has increased to 55%. Note here that the surface energy of the crystal facets varies with: A (101) < R (110) < A (100) < R (100) < R (001)[36]. It is therefore likely that the rutile crystals having (110) orientation are the most stable, as the system is the most stable when least energetic facets are interfacing with vacuum (air) or the substrate. They are likely to become the nucleation seeds for the formation of mono-oriented rutile platelets (110). The growth of these platelets can occur by a stochastic reorganization of the material at the interface with both the anatase and the rutile particles. At 30 s, the rutile phase takes over the entire crystallinity, divided between areas of poly-oriented NPs and areas of mono-oriented platelets (110). Thus, it appears that platelet growth is dominated by the consumption of anatase NPs, since it is more favourable than both the transformation of anatase NPs into rutile NPs, as well as the sintering of rutile NPs into platelets when the latter do not exhibit the (110) orientation. The system still has the freedom to evolve into an assembly of platelets grown by diffuse sintering with the remaining poly-oriented rutile NPs between 30 and 60 s.

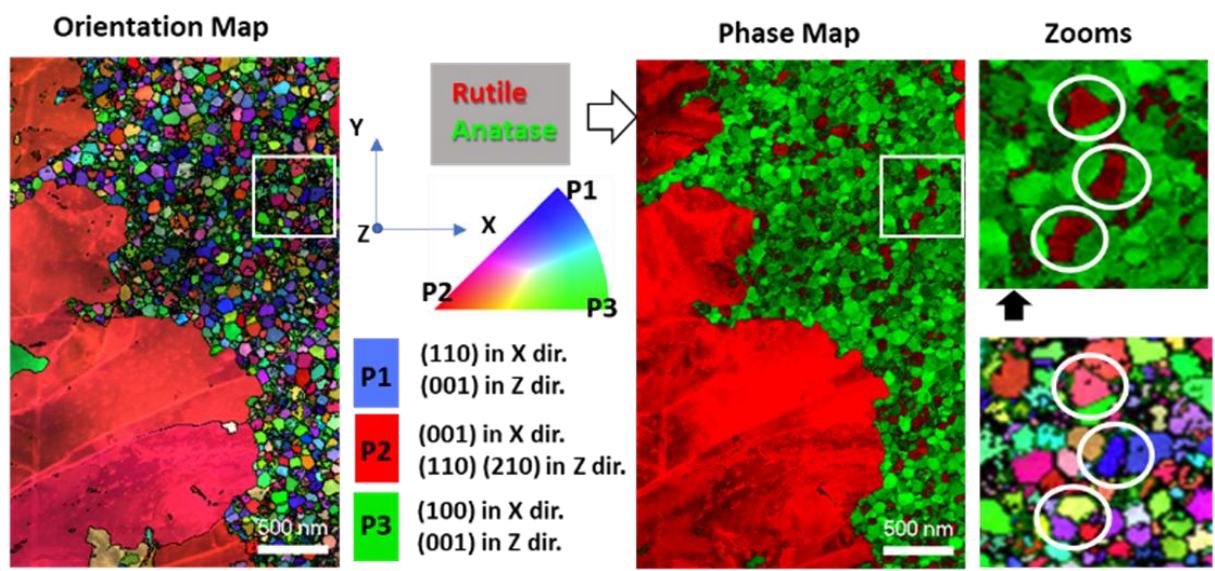

Figure 3. PED-TEM analysis of the coating treated 20 s at 1100 °C.

Concerning the optical properties, the index increases, as expected, with the rutile content[17]. However, this cannot occur without the formation of large platelets (as they grow by consuming the anatase NPs). Since these platelets are responsible for light scattering (the index and diffusion increase considerably when the proportion of rutile increases from 55 to 100% between 20 and 30 s), it turns out that the formation of a rutile film with similar optical quality to that of the anatase film is difficult. Nevertheless, the index and scattering values obtained for anatase -rich samples obtained with lower thermal budget meet high-performance

| | | | RI@520nm | S@520nm (Sub. Cor.) |
|---|---|---|---|---|
| 5s | 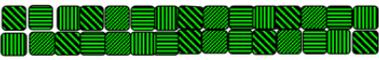 | Anatase poly-oriented NP | 2.508 | 0.48 % |
| 10s | 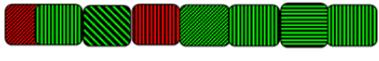 | Anatase and Rutile poly-oriented NP | 2.513 | 0.58 % |
| 20s | 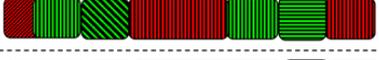 | Anatase and Rutile poly-orientated NP + (110) Rutile platelets | 2.563 | 0.62 % |
| 30s | 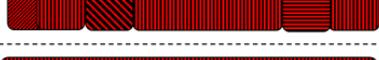 | Rutile poly-orientated NP + (110) Rutile platelets | 2.698 | 0.94 % |
| 60s | 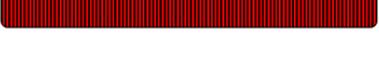 | (110) Rutile platelets | 2.726 | 1.17 % |

Figure 4. Scheme illustrating the mechanism of rutile platelets formation from sol-gel formulation with time at 1100°C and the corresponding optical properties.

standards for optical and photonic applications.

## 4. Conclusions

In conclusion, the formation of high index TiO$_2$ thin films (n = 2.7) is a complex mechanism involving nucleation growth, phase transformation and selective sintering into rutile platelets mono-oriented with respect to the amorphous substrate surface. Experimental data reveal that this transformation requires annealing for a few tens of seconds at 1100 °C. rutile formation occurs through two mechanisms: (i) the transition from the (112) anatase facets into the (100) rutile facets while keeping the initial orientation, and (ii) the lateral growth of the (110) oriented rutile crystals by consumption of the adjacent anatase nanocrystals (the (110) rutile face being the least energetic). An intermediate stage exists, in which the film consists of micron-sized areas of poly-oriented rutile nanocrystals and mono-oriented rutile platelets (110), that evolves by diffusive sintering towards the final stage, where the film is made up solely of mono-oriented platelets. It thus appears that the formation of mono-oriented rutile platelets by sintering with adjacent anatase particles is more favourable than the transformation of isolated anatase NPs into rutile, suggesting that the formation of a 100% rutile film without platelets is difficult under these conditions. This transformation is accompanied by an increase in the refractive index of $TiO_2$ up to 2.72 (at 520 nm), and an increase in scattering from 0.5 to 1.17%, associated with the formation of platelets. Accessing low-cost processing of high-refractive index coatings is still in high demand to consider replacing electronic circuitry. Using sol-gel deposition techniques, it was possible to overcome major processing limitation and fabricate thin films with great potential for optical and photonic applications requiring extreme indices and relatively low scattering.

### 5. Acknowledgements

The authors are thankful for the support of the IUF